# A Bayesian Updating Scheme for Pandemics: Estimating the Infection Dynamics of COVID-19

Shuo Wang#, Xian Yang#, Ling Li, Philip Nadler, Rossella Arcucci, Yuan Huang, Zhongzhao Teng, Yike Guo*

*Abstract*—Epidemic models play a key role in understanding and responding to the emerging COVID-19 pandemic. Widely used compartmental models are static and are of limited use to evaluate intervention strategies with the emerging pandemic. Applying the technology of data assimilation, we propose a Bayesian updating approach for estimating epidemiological parameters using observable information for the purpose of assessing the impacts of different intervention strategies. We adopt a concise renewal model and propose new parameters by disentangling the reduction of instantaneous reproduction number $R_t$ into mitigation and suppression factors for quantifying intervention impacts at a finer granularity. Then we developed a data assimilation framework for estimating these parameters including constructing an observation function and developing a Bayesian updating scheme. A statistical analysis framework is then built to quantify the impact of intervention strategies by monitoring the evolution of these estimated parameters. By Investigating the impacts of intervention measures of European countries, the United States and Wuhan with the framework, we reveal the effects of interventions in these countries and the resurgence risk in the USA.

## I. Introduction

In response to the COVID-19 pandemic, governments have taken non-pharmaceutical intervention measures. Common measures include travel restriction, school and non-essential business closure and social distancing, as well as early isolation of confirmed patients. Recently, as the first-wave epidemic peak has faded away in many countries, the accumulated observations of epidemic growth [1] and corresponding intervention policies [2] shed more insights on how the interventions worked. Meanwhile, many governments have switched into the phase to reopen economic and social activities, with attention on tamping down possible resurgences. However, the recent second-wave outbreak in some countries and regions (e.g. the United States, Hong Kong) alerts us to monitor the epidemic evolution carefully while intervention measures are being relaxed.

Mathematical models play a key role in understanding and responding to the emerging COVID-19 pandemic [3]–[5]. Compartmental models (e.g. SIR, SIER) and time-since-infection models (i.e. renewal process-based models) are the two well-known approaches describing the underlying transmission dynamics [6], [7]. The compartmental models describe the transmission among sub-populations while the renewal process-based approach starts from the inter-individual transmission. Despite different nomenclatures and applications, each model contains parameters characterizing the epidemic dynamics. One of the most well-known parameters is the reproduction number $R$, which represents the average number of secondary cases that would be induced by an infected primary case [8]. This key parameter is related to the final epidemic size of infectious disease [9]. Intervention measures aim to maintain the reproduction number under one so that the epidemic can be contained along with time. Thus, estimation of time-varying $R$ will reflect the impacts of intervention.

The basic reproduction number $R_0$ is the reproduction number at the beginning of the epidemic outbreak, when the susceptible population is approximately infinite and without intervention measures. When various intervention measures are being introduced, the instantaneous reproduction number $R_t$ (also called effective reproduction number) is of greater interest. To gain insights into epidemic evolution, most existing studies such as [3], [10] focus on estimating time-varying instantaneous reproduction number $R_t$. $R_t$ is defined as the average number of secondary cases that would be generated by an infected primary case at a time $t$ when conditions remained the same thereafter [8], reflecting the real-time transmission dynamics. This could help governments to monitor the evolution of COVID-19 and update intervention policies accordingly [11].

However, the nowcasting of $R_t$ from reported data is not an easy task. Several approaches have been proposed to estimate $R_t$ with different advantages [12]–[14], but the timeliness and accuracy are still of concern. Nowcasting results are affected by different factors, such as assumptions of the epidemic models, statistical inference methods and uncertainty of data resources. Inappropriate interpretation or imprecise estimation of $R_t$ are criticized for providing misleading information [15]. For example, the nowcasting from reported confirmed cases will fall behind the nowcasting from onset data because there is a delay from symptom onset to case report. We hypothesize that more detailed characteristics of the time-varying infectiousness profile could be estimated from the publicly available reports (e.g., death data, confirmed data, onset data and laboratory data) and help better understand and evaluate the efficiency of interventions.

* Corresponding author: Yike Guo (email: yikeguo@hkbu.edu.hk).
# Shuo Wang and Xian Yang contribute equally.
Shuo Wang, Philip Nadler and Rossella Arcucci are with the Data Science Institute, Imperial College London, London, UK.
Xian Yang is with the Department of Computer Science, Hong Kong Baptist University, Hong Kong, China.
Ling Li is with the School of Computing, University of Kent, Kent, UK.
Yuan Huang and Zhongzhao Teng are with the Department of Pure Mathematics and Mathematical Statistics, University of Cambridge, Cambridge, UK.
Yike Guo is with Hong Kong Baptist University, Hong Kong, China.

In this study, we propose a comprehensive Bayesian updating scheme for reliable and timely estimation of parameters in epidemic models. The transmission dynamics are modelled as a concise renewal process with time-varying parameters. To monitor the evolving impacts, more fine-grained modelling of the transmission dynamics is required. Instead of the well-known $R_t$, we introduce two complementary parameters, the mitigation factor ($p_t$) captures the effect of shielding susceptible population (e.g. through social distancing), and the suppression factor ($D_t$) captures the effect of isolating the infected population (e.g. through quarantine) to stop virus transmission. We propose a novel method to estimate these parameters by taking the data assimilation approach of using Bayesian updating methods. We use daily reports of confirmed cases as the observation. A deconvolution method is used to build an observation function to estimate the infection cases by adjusting the incubation time and report delay. The evolution of the time-varying infectiousness profile (i.e. $p_t$ and $D_t$) is estimated from the adjusted epidemic curve through a Bayesian approach of assimilation. Such a fine-grained infectiousness profile enables us to quantify the impacts of various intervention measures in a comprehensive way.

The paper is structured as follows: We introduce the related work in Section II. In section III, we present the overview of a time-varying renewal process model where the two parameters $p_t$ and $D_t$ are proposed. In section IV, we present in detail the Bayesian updating scheme for estimating the dynamic parameters. In section V, we develop a statistical analysis method of assessing the intervention impacts based on the estimated results and the report of intervention policies. In section VI, as applications of our approach, we investigate the impacts of intervention measures of European countries, the United States and Wuhan to illustrate the importance of this development.

## II. RELATED WORK

At the beginning of COVID-19 outbreak in Wuhan, China, compartmental models (e.g. SIR, SEIR model) have been used to investigate the epidemic dynamics [16]–[18], where the basic reproductive number was estimated from the models with static parameters. With the spread of COVID-19 worldwide, renewal process-based models (i.e. time-since-infection model) are also being widely used in the study of COVID-19. The R package 'EpiEstim' [12], [13] is the most widely used in estimating the time-varying $R_t$ with a sliding window. In [10], 'EpiEstim' was applied to infer $R_t$ via the discrete renewal process for policy impact assessment. Similar work has been done in [3] to infer $R_t$ using 'EpiEstim' from laboratory-confirmed cases in Wuhan and hence evaluated the impact of non-pharmaceutical public health interventions. The work in [11] has pointed out that the infection data is usually not available and death data was used as observation for $R_t$ updating. Instead of simply applying 'EpiEstim', they estimated $R_t$ by employing the renewal equation as a latent process to model infections and connecting the infections to death data via a generative mechanism. However, the estimated $R_t$ is in a piecewise form and the number of changing points was assumed to be determined by the imposed interventions. [19] estimates $R_t$ from the death data as well while linking the disease transmissibility to mobility using the renewal equation. In general, [11] and [19] explicitly formulated the $R_t$'s updating function by introducing external factors (e.g. interventions and mobility). Thus, the estimated $R_t$ curve is largely constrained by the factors that are considered in the model.

Data Assimilation [20] lends itself naturally to this problem since it provides a framework to enable dynamically updating the model states and parameters when new observations become available while also taking into account model and observation uncertainty. Data assimilation technologies, such as Kalman filter and variational method [21], have been widely used in signal tracking, oceanology, environment monitoring and weather forecasting where physical models and observation data are assimilated to produce accurate prediction. Data assimilation for epidemiological modelling was first proposed in [22] where compartment models were used as the underlying model for assimilation. In [25] and [26], estimating time-varying parameters in the compartment models was further investigated. To the authors' best knowledge, our work is the first study of applying data assimilation to the renewal process-based model.

## III. EPIDEMIC MODELLING OF COVID-19 TRANSMISSION

In this section, we propose a time-varying renewal process with two complementary parameters $p_t$ and $D_t$ to model the evolving infectiousness profile. We adopted a time-varying renewal process for epidemic modeling. The renewal process [8] of infectious disease transmission is:

$$I(t) = \int_0^\infty I(t-\tau)\beta(\tau)d\tau \qquad (1)$$

where $I(t)$ is the incident infection on time $t$ and $\beta(\tau)$ is the infectiousness profile. The infectiousness profile means a primary case who was infected $\tau$ time ago (i.e. with the infection-age $\tau$) can now generate new secondary cases at a rate of $\beta(\tau)$, describing a homogenous mixing process. The infectiousness profile $\beta(\tau)$ is related to biological, behavioral and environmental factors. We can calculate the reproduction number $R$ as the area under curve of $\beta(\tau)$, which is the overall number of secondary cases infected by a primary case. Further, $\beta(\tau)$ can be rewritten as:

$$\beta(\tau) = R \cdot w(\tau) \qquad (2)$$

where the unit-normalized transmission rate $w(\tau)$ is the probability density function of generation time, i.e. the interval between the primary infection and the secondary infection. In the early stage without intervention, the infectiousness profile remains time-independent as the baseline $\beta_0(\tau)$ which describes the transmission dynamics when the susceptible population is infinite. The corresponding $R$ is the well-known basic reproduction number $R_0$. In reality, the infectiousness profile $\beta(\tau)$ will evolve with time $t$, therefore we introduce $\beta_t(\tau)$ to address the change in its distribution caused by intervention measures.

To quantify the impacts of intervention measures to the evolution of $R_t$, we propose two factors: **suppression** and **mitigation** to disentangle the intervention effects. Here we use two complementary metrics $p_t$ and $D_t$ modelling the suppression and mitigation factors respectively, as illustrated in Figure 1.

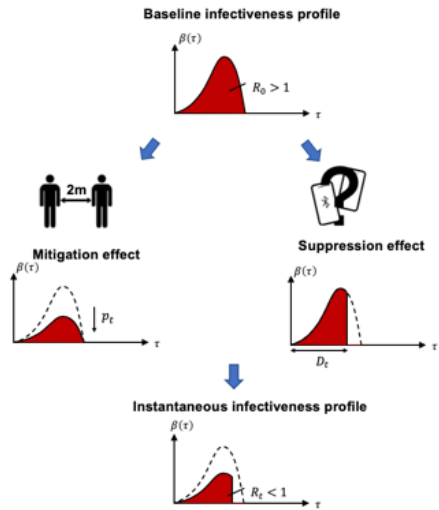

Fig. 1. Disentangling the reduction of reproduction number into mitigation and suppression factors.

The suppression effects mainly shorten the infectious period of the infected population, corresponding to the truncation of $\beta(\tau)$ along the horizontal axis. We use a time-varying parameter $D_t$ to denote the effective infectious window induced by suppression. The mitigation effects attenuate the overall infectiousness by shielding the susceptible population, corresponding to the scaling on the vertical direction. We introduce another time-varying parameter $p_t$ to describe this attenuation effect induced by mitigation. Formally, we parameterize the evolution of the infectiousness profile as:

$$\beta_t(\tau) = \begin{cases} \beta_0(\tau) \cdot p_t & \tau < D_t \\ 0 & \tau \geq D_t \end{cases} \quad (3)$$

Accordingly, the instantaneous reproductive number $R_t$ can be derived:

$$R_t = p_t \cdot \int_0^{D_t} \beta_0(\tau) d\tau \quad (4)$$

Therefore, the impact of intervention measures on $R_t$ reduction is disentangled: mitigation factor $p_t$ attenuates the overall infectiousness through shielding the susceptible population and suppression factor $D_t$ shortens the infectious period through isolating the infected population. It is noted that the $R_t$ can be derived from $p_t$ and $D_t$ which provide more mechanistic details about the evolution of the infectiousness profile.

IV. ADAPTIVE PARAMETER ESTIMATION

We aim to develop a comprehensive framework to estimate parameters of renewal process models using Bayesian updating approach of data assimilation, especially the three key parameters: $<R_t, p_t, D_t>$. The estimation is essential for quantify the impacts of different interventions through monitoring the evolution of $<R_t, p_t, D_t>$. This framework contains building an observation function to map observations to model state, modelling and Bayesian updating as shown in Figure 2 and 3. By applying the observation function, we reconstruct the number of daily infections from reports of confirmed cases, taking into account the incubation time and report delay with a deconvolution algorithm. Then $<R_t, p_t, D_t>$ is estimated through a Bayesian approach of data assimilation.

*A. Reconstruction of daily infection from reported cases*

In data assimilation, model states and parameters can be updated using new observation data. It is important for parameter estimation that proper observation is chosen, and an observation function can be built which maps observations to a state variable (usually regarded as the output of the model).

In this study, the observations we have chosen are from the reported number of confirmed cases. The model output is daily infection incidence through the renewal process. However, such observations experience an inevitable time delay between the actual infection time and the reporting date (Figure 2). This includes an incubation time (i.e. the period between infection and onset of symptoms) and confirmation period (i.e. the period between onset and officially reported after being tested). The confirmed cases reported on time $t$ were actually infected within a past period and the reported number is the convolution result of the historical daily infection.

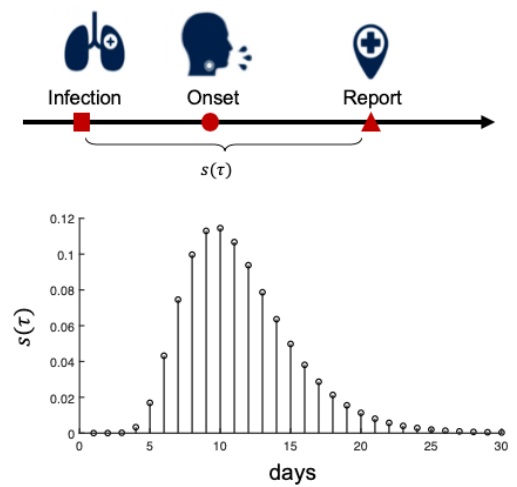

Fig. 2. Reconstruction of daily infection from the confirmed cases using deconvolution algorithms. The time delay between the infection and onset and report is demonstrated (top). The estimated distribution between infection and report is presented which is used for deconvolution (bottom).

Here, we define an observation function to reconstruct the daily infection instances from the confirmed cases using the deconvolution technique with Richardson-Lucy (RL) iteration method [25]. We use the incubation period calculated by Ferretti et al.[5], which is a lognormal distribution with a mean of 5.5 days and a standard deviation of 2.1 days. We use the confirmation period previously reported by Leung et al. [10], which is a gamma distribution with a mean of 4.9 days and a standard deviation of 3.3 days. Sampling from these two sequential distributions, we estimated the discrete interval distribution $s(\tau)$ for $\tau \in \{0, d\}$ from infection to report (Figure 2). Denoting the epidemic curve of reported infection cases $\hat{I}_{1:t} = \{\hat{I}_1, \hat{I}_2, \ldots, \hat{I}_t\}$ and the epidemic curve of confirmed cases $C_{1:t} = \{C_1, C_2, \ldots, C_t\}$, the reported infection with an observation process of past infections can be modelled as a Poisson process:

$$C_t \sim Poisson(mean = \sum_{k<t} s(k)\hat{I}_{t-k}) \quad (5)$$

Estimate the daily reported infection curve $\hat{I}_{1:t}$ given the daily confirmed cases curve $C_{1:t}$ and infection-to-confirmed time distribution $s_{1:d}$ is an ill-posed deconvolution problem and can be solved using Richardson-Lucy (RL) iteration method [25]. The initial guess $\hat{I}^0_{1:t}$ is the confirmed cases curve $C_{1:t}$ shifted back by the mode of the infection-to-confirmed time distribution. Let $\hat{C}^n_i = \sum_{k<t} s(k)\hat{I}^n_{t-k}$ be the expected number of confirmed cases on day $i$ of iteration $n$, and $q_t$ be the probability that a reported case resulting from infection on day $t$ will be observed as defined in [25]. Then the iteration of $\hat{I}_t$ is computed by an expectation-maximization (EM) algorithm as:

$$\hat{I}^{n+1}_t = \frac{\hat{I}^n_t}{q_t} \sum_{i>t} \frac{s(i-t)C_t}{\hat{C}^n_t} \quad (6)$$

A normalized $\chi^2$ statistics is used as the stop criterion of the iteration:

$$\chi^2 = \frac{1}{N} \sum_t \frac{(\hat{C}^n_i - C_t)}{\hat{C}^n_i} < 1 \quad (7)$$

where $N$ is the total number of data points. It is of note that the reported number of confirmed cases constitute the lower bound of the real infection due to the lack of mass test and the existence of asymptomatic cases. However, as long as the detection rate remains consistent, the scaling of reconstructed data does not affect the following inference of transmission dynamics.

### B. Bayesian Updating for Parameter Estimation

Following the Bayesian updating approach of data assimilation, we propose an instantaneous estimation method. For the defined epidemiology renewal process, the daily incident infection $I_t$ is the state variable and can be assimilated from the reconstructed infection data from observation. The evolution of the state $I_t$ is governed by the renewal process with the time-varying infectiousness profile $\beta_t(\tau)$, parameterized with $p_t$ and $D_t$. Here we present a Bayesian framework to monitor the evolution of $p_t$ and $D_t$ using the daily reports of confirmed cases (Figure 3).

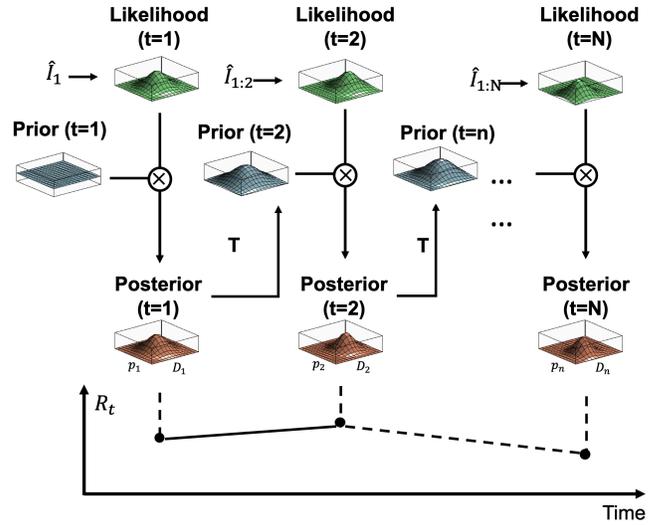

Fig. 3. Illustration of the Bayesian updating framework for estimating suppression and mitigation factors. We employ a two-level hierarchical model: For each time step, the low-level model (i.e. renewal process) provides the likelihood of $p_t$, $D_t$ (green). The posterior (orange) is calculated through the element product of the likelihood and the prior (blue) from the previous time step. To generate the prior for next time step, we use the high-level model (i.e. the transformation T) to induce the evolution of parameters. The high-level model is a piecewise gaussian random walk process where the fluctuations of $p_t$ and $D_t$ differ before and after an intervention time. The instantaneous reproduction number $R_t$ can be derived from the posterior distribution of $p_t$ and $D_t$.

Our updating scheme employs a two-level hierarchical model for the inference of time-varying parameters [26]. Let us denote the observed daily incidence of infection till time step $t$ as $\hat{I}_{1:t} = \{\hat{I}_1, \hat{I}_2, \ldots, \hat{I}_t\}$. Suppose $p(\theta_{t-1}|\hat{I}_{1:t-1})$ is the estimated distribution of $\theta = [p, D]^T$ at time step $t-1$. Under the assumption of consistent detection rates, the observed daily incidence $\hat{I}_t$ also satisfies the renewal process. The low-level model predicts the observation (i.e. reconstructed daily infection) given a parameter set through the renewal process:

$$p(\hat{I}_t|\theta_t, \hat{I}_{1:t-1}) \sim Poisson(mean = \sum_{k=1}^{t-1} \beta_t(k; \theta_t)\hat{I}_{t-k}) \quad (8)$$

where a Poisson process of observing the infected cases is assumed. This describes the likelihood of observing the new incidence data given history observations and parameter value $\theta_t$. The high-level model describes the evolution of the model parameters $p_t$ and $D_t$ through transforming the joint distribution:

$$p(\theta_t|\hat{I}_{1:t-1}) = T \circ p(\theta_{t-1}|\hat{I}_{1:t-1}) \quad (9)$$

where $T(.)$ is a transformation function defining the temporal variations of the $\theta$. The prior knowledge of parameter distribution is transferred to the next time step $t$ by the high-level model T. Under the scenario without interventions, the parameters $p_t$ and $D_t$ fluctuate around the baseline values. Therefore, we can assume a random walk of $\theta$ in the parameter space as the high-level model. The update of joint parameter distribution is by convoluting with a Gaussian kernel with variance $\sigma_1$. When the intervention is introduced on time $d$, the random walk of $\theta$ is altered where the variance of the Gaussian kernel will become $\sigma_2$. The transformation $T(.)$ is defined as:

$$T \circ p(\boldsymbol{\theta}) = \begin{cases} p(\boldsymbol{\theta}) * K_{\sigma_1}(\boldsymbol{\theta}) & t < d \\ p(\boldsymbol{\theta}) * K_{\sigma_2}(\boldsymbol{\theta}) & t \geq d \end{cases} \quad (10)$$

where $K_{\sigma_1}(\boldsymbol{\theta})$ and $K_{\sigma_2}(\boldsymbol{\theta})$ are the Gaussian kernels before and after the deployment of intervention at time $d$. This high-level model includes three hyperparameters: variances before and after intervention: $\sigma_1$ and $\sigma_2$, and the change-point time $d$. Let us denote the hyperparameters $\boldsymbol{\eta} = [\sigma_1, \sigma_2, d]^T$. After seen the latest observation $\hat{I}_t$, the posterior estimation of $\boldsymbol{\theta}$ is update by the Bayes rule:

$$p(\boldsymbol{\theta}_t|\hat{I}_{1:t}) = \frac{T \circ p(\boldsymbol{\theta}_{t-1}|\hat{I}_{1:t-1}) \cdot p(\hat{I}_t|\boldsymbol{\theta}_t, \hat{I}_{1:t-1})}{p(\hat{I}_t|\hat{I}_{1:t-1})} \quad (11)$$

This step reflects the Bayesian principle in the key updating step in Kalman filtering [21]. Unlike the Kalman filtering method where uncertainty is explicitly modelled through a covariance matrix under the Gaussian assumption, we directly use posterior probability to capture the uncertainty of estimation. The posterior is usually intractable but can be approximated through grid-based methods. Given a set of hyperparameters $\boldsymbol{\eta}_i$, the hybrid model evidence can be calculated as [26]:

$$p(\hat{I}_{1:t}|\boldsymbol{\eta}_i) = \int p(\hat{I}_{1:t}, \boldsymbol{\theta}_t|\boldsymbol{\eta}_i) d\boldsymbol{\theta}_t \quad (12)$$

Finally, the posterior estimation $p(\boldsymbol{\theta}_t|\hat{I}_{1:t})$ can be averaged across the hyperparameter grids weighted by the hybrid model evidence. The posterior mean and confidence intervals of $p_t$ and $D_t$ as well as the corresponding $R_t$ are obtained in a dynamic manner. The prior of $R_0$ at the first timestep is set uninformative as a uniform distribution with the pre-set lower and upper limits (e.g., the upper limit for the European countries is set to 8 in the experiment). The shape of $\beta_0(\tau)$ is adapted from the distribution of generation time interval $w(\tau)$ reported by Ferretti et al.[5] We applied the above framework to infer the epidemic evolution in 14 European countries, states in the US and Wuhan city, China in Section VI. The codes of the our framework is released as an open-source package (https://github.com/whfairy2007/COVID19_Bayesian).

## V. EVALUATION OF INTERVENTION MEASURES

With the estimated results from the above Bayesian updating scheme, now we can perform statistical analysis between the evolution of the transmission dynamics and the implementation of intervention measures. The whole framework containing data reconstruction, dynamic modelling, Bayesian updating, statistical analysis is presented in Figure 4. In this section, we introduce the quantification of intervention measures and the statistical method.

### A. Data Source

For the observations, we use the aggregated data of publicly available daily confirmed cases of 14 Europe countries (Austria, Belgium, Denmark, France, Germany, Ireland, Italy, Netherlands, Norway, Portugal, Spain, Sweden, Switzerland and the United Kingdom) and 52 states of the United States from John Hopkins University database [1]. The data include the time series of confirmed cases from January 22nd to June 8th 2020 (accessed on June 9th 2020). Six states with accumulated confirmed cases less than 1,000 are excluded from the analysis. The daily number of onset patients in Wuhan is adopted from the retrospective study by Pan et al. [3].

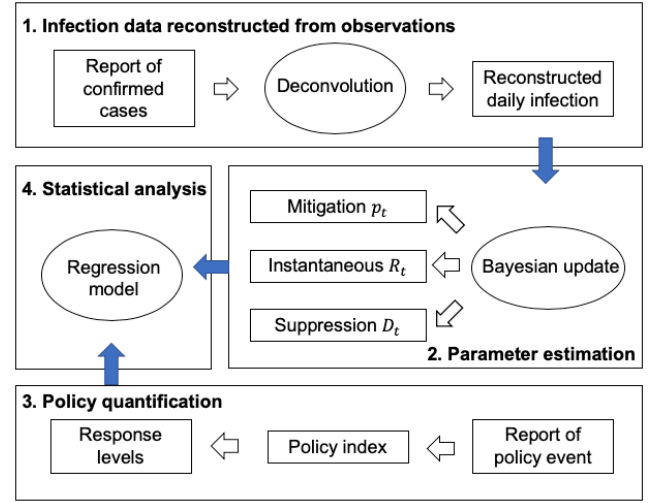

Fig. 4. Components of the quantification framework. The evolution of mitigation and suppression factors are estimated using the infection data reconstructed from the daily reported confirmed cases. Given the history of government responses, the impacts of intervention measures are quantified by correlating the inferred epidemic parameters to response levels.

The data of intervention measures in European countries are collected from the Oxford Coronavirus Government Response Tracker [2], reporting the overall stringency index $S_t$ of intervention measures during the analysis period (accessed on June 9th 2020). This overall stringency index is calculated based on the policy quantification of eight intervention measures (i.e. School closing, Workplace closing, Cancel public events, Restrictions on gatherings, Close public transport, Stay-at-home requirements, Restrictions on internal movement and International travel controls) and one health measure (i.e. public info campaigns) to indicate the government response level of intervention.

According to the normalized stringency index by Oxford report [2], we categorized the dates into five response levels (Level 0: $S_t \leq 20\%$, minimal response for reference; Level 1: $20\% < S_t \leq 40\%$, soft response; Level 2: $40\% < S_t \leq 60\%$, strong response; Level 3: $60\% < S_t \leq 80\%$ and Level 4: $80\% < S_t \leq 100\%$, emergent responses). The representative intervention measures for each response level were identified based on the contribution to the stringency index $S_t$.

### B. Calculation of intervention policy indices

We categorize the dates within our analysis period in European countries into five different response levels, based on the overall stringency index $S_t$. To identify the representative measures of each response level, we calculate the quantification indices of the eight intervention measures. Descriptions of the eight intervention measures and the quantification methods are provided in [2]. For each intervention measure, the Oxford report provides an ordinal scale quantification $v_{j,t}$ of the strength of j-th policy implementation and a binary flag $f_{j,t}$ representing whether it is implemented in the whole country on time $t$. Following similar practice use in the Oxford report, we normalize the implementation of each intervention measure as

$$P_{j,t} = \frac{\max(0, v_{j,t} + 0.5 f_{j,t} - 0.5)}{N_j} \times 100\% \quad (13)$$

where $N_j$ is the maximum value of the indicator $P_j$. To assign a label of response level to each measure, we calculate the change of mean policy indices across different response levels. The response level with the largest increase is considered as the level that the measure belongs to (i.e. the measure is a representative measure of this response level). For example, the mean index of school closure showed the largest increase from Level 0 to Level 1, so we consider this is a representative measure of Level 1. The representative measures of each response level are listed in Table 1.

### C. Regression analysis of the intervention impacts

We performed a retrospective analysis of the time-varying transmission dynamics during different response levels in Europe countries. First, the evolution history of $R_t$ and the overall stringency index $S_t$ are obtained using the above framework. The stringency index $S_t$ is categorized into five response levels. We fit a log-linear mixed-effect model, where the logarithm of $R_t$ is the outcome variable and categorical stringency index is the predictor. The logarithm is used to obtain the intervention impacts on the relative change of $R_t$ [27]. We performed a partial-pool analysis by assuming the impacts of intervention measure (slopes) shared across all selected European countries while the basic reproduction number $R_0$ (intercept) varies due to environmental and social factors. The regression formula is written as:

$$\ln R_{j,t} = b_0 + \sum_{k=1}^{4} b_k * D_{j,k} + \gamma_j + \epsilon \quad j = 1,2,\dots,14 \quad (14)$$

where $R_{j,t}$ is the estimated reproduction number of j-th country, $b_0$ is the fixed effect term of $\ln R_0$ and $b_k$ is the fixed effects of interventions in response level $k$. $D_{j,k}$ is the dummy variable that takes the value 1 if and only if the response status is at Level k. $\gamma_j$ is the random effect term following zero-mean Gaussian which explains the difference of $\ln R_0$ across countries and $\epsilon$ is the Gaussian error term. Equation 14 associates the relative changes in $R$ to the fixed effects of response levels, and can be rewritten into its marginal form as:

$$\ln(1 + \frac{R - R_0}{R_0}) = \sum_{k=1}^{4} b_k * D_k \quad (15)$$

Therefore, the relative change of $R$ due to the intervention measures in $k$-th response level can be derived from $b_k$ (i.e. $\Delta R/R_0 = exp(b_k) - 1$). Country-specific $\ln R_0$ can be estimated as $b_0 + \gamma_j$ at the Level 0. The statistical analysis is performed using the R package 'lme4'. The fixed effect is considered significant with P value<0.05. The 95% confidence intervals (CI) are estimated using bootstrap method. The assumption of normality is checked by inspecting the quantile-quantile plot of the residuals. The same procedure is also applied to the analysis of $D_t$ and $p_t$ to quantify the suppression and mitigation factors, respectively. The results are demonstrated in Table 1.

## VI. RESULTS

### A. Validation on simulated data

We simulated an artificial epidemic outbreak with a time-varying infectiousness profile using renewal process. The generation time intervals were adapted from Ferretti et al.[5]. The simulation period includes 50 days and an intensive intervention measure is induced on day 35 altering the transmission dynamics. Before the intervention, the ground-truth $R_t$ followed Gaussian random walk with a mean of 2.5. After the intervention (50% $p_t$ reduction and 67% $D_t$ reduction), the mean of $R_t$ was reduced to 0.5 (black line).

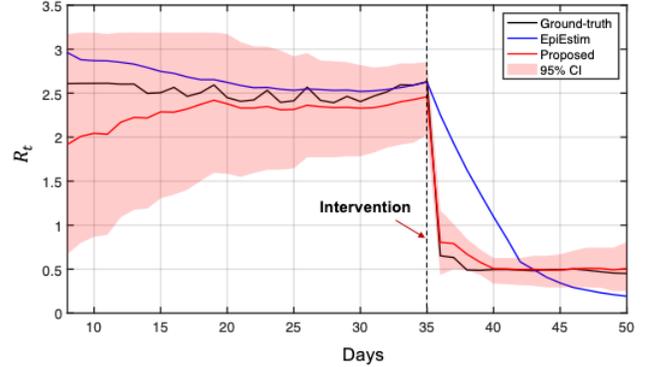

Fig. 5. Validation of the proposed Bayesian updating scheme.

We validate the effectiveness of our approach in capturing the sudden change of $R_t$ evolution induced by interventions, which is hard to be detected by traditional sliding window-based methods (Figure 5). We compared the results using our approach (red line with 95% confidence intervals) to the results computed by the R package 'EpiEstim v2.2' [12] (blue) which is a sliding window-based method widely used for $R_t$ estimation. We observed that the ground-truth $R_t$ is well estimated within our confidence interval. In particular, the sharp change of $R_t$ caused by the intervention is captured immediately by our approach while there is a lag using the sliding window-based method.

### B. Evaluation of Intervention measures in Europe Countries

In this part, we applied the proposed framework to analyze the epidemic evolution in the 14 European Countries and also Wuhan. With the inferred <$R_t, p_t, D_t$>, we can then assess the impacts of intervention measures.

Figure 6 demonstrates the reconstruction of daily infections in the UK from the reported confirmed cases. The infected-to-report delay between report and infected time is composed of the incubation period (a lognormal distribution with a mean of 5.5 days and a standard deviation of 2.1 days [5]) and the onset-to-report period (a gamma distribution with a mean of 4.9 days and a standard deviation of 3.3 days [10]). The blue bars in Figure 6 indicate the number of confirmed cases. After deconvolving the confirmed numbers using infected-to-report delay, we got the infected curve, which is colored in red in

Figure 6. To check the reliability of the deconvolution results, we convolve the inferred infected curve (in red) with the infected-to-report delay to recover the confirmed curve (in black). We can see that the black curve matches well to the original blue bars and is much smoother. With the above observation, we can see the effectiveness of the infected curve inference. Figure 7 shows the results of estimating $R_t$ of the UK from the infected curve. The missing values in the infected curve are replaced by the average mean of the neighbouring numbers. green bar is the posterior mean of estimated $R_t$.

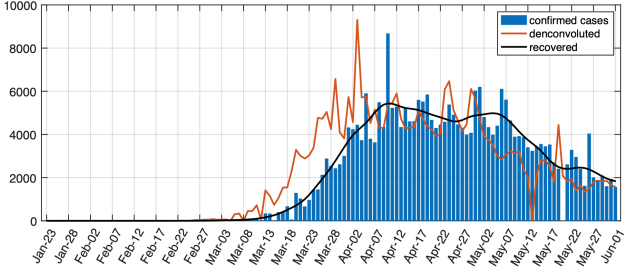

Fig. 6. Reconstruction of daily infections from the report of confirmed cases in UK. The forward convolution on reconstructed data (black line) matches well with actual reported data (blue bars), validating the correctness of the deconvolution method.

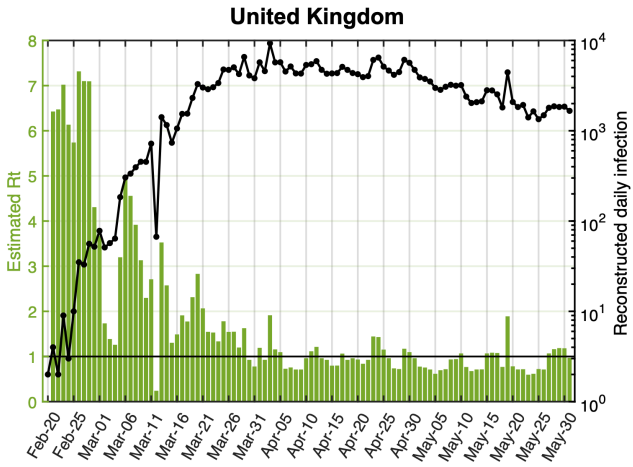

Fig. 7. Estimated evolution of transmission dynamics in UK. The black line represents the reconstructed daily infection number and the green bar is the posterior mean of estimated $R_t$.

To quantitatively show the impacts of different strength levels of interventions, Table 1 summarizes the statistical analysis results of 14 European countries. It shows different reduction rates of $<R_t, p_t, D_t>$ for different response levels. The relative reduction of $<R_t, p_t, D_t>$ compared to the minimal response (Level 0 where $R_t$ is set to $R_0$) was estimated for each response level. With soft response (Level 1), the corresponding intervention measures (e.g. school closure, quarantine of international arrivals from high-risk regions) are correlated with a relative reduction of $R_t$ by 35% showing both strong suppression effect ($D_t$ shortening 22%) and mitigation effect ($p_t$ reduction 29%). With strong response (Level 2), the relative reduction of $R_t$ increases to 60% with a strong mitigation effect ($p_t$ reduction 56%). But the suppression effect ($D_t$ shortening 26%) is similar to that of Level 1, indicating marginal incremental suppression effect. This observation shows a consistency with the aim of representative intervention measures on this level (e.g. cancelling public events, restrictions on gathering and internal movements) to reduce the contact rates among the population.

The emergent response (Level 3) shows substantial relative reduction of reproductive number ($R_t$ reduction 71%) with suppression ($D_t$ shortening 37%) and mitigation ($p_t$ reduction 67%) effects, correlated to the intensive measures (e.g. workplace closure and stay-at-home requirements). A similar degree of reductions is found for Level 4 ($R_t$ reduction 74%; $D_t$ shortening 40%; $p_t$ reduction 70%) while the stringency of intervention measures is higher. We find that our estimated evolving patterns of $p_t$ and $D_t$ correspond well to the serial strategies taken by some European countries, such as the 'contain-delay-lockdown' route taken in the UK.

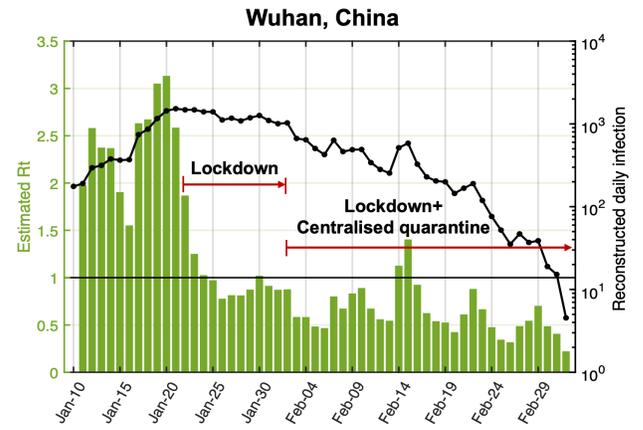

Fig. 8. Estimated evolution of transmission dynamics in Wuhan. The black line represents the reconstructed daily infection number and the green bar is the posterior mean of estimated $R_t$. Two major events (city lockdown measure from Jan 23rd and centralized quarantine from Feb 2nd) are annotated with red arrows.

Apart from the results of 14 European Countries, Figure 8 also shows the results of applying our method to the data from Wuhan, where the greens bars indicate the posterior mean of $R_t$ during the outbreak of COVID-19. We can see that at the early stage of the pandemic, the $R_t$ levels are above 1. After the lockdown intervention has taken effect, $R_t$ has experienced a sharp decrease from 23rd Jan. When the centralized quarantine policy has been enforced from the beginning of February, the $R_t$ values then largely remain below zero (the spike around 14th Feb is due to misreporting).

Figure 9 compares the reductions in $<R_t, p_t, D_t>$ for different response levels between European Countries and Wuhan. From the analysis of Wuhan data, the strong impact of lockdown is clearly demonstrated with the immediate relative reduction of $R_t$ by 58%. We also observed that the combination of lockdown, centralized quarantine and immediate admission of confirmed patients starting from Feb 2nd in Wuhan was associated with a more substantial relative reduction of $R_t$ with strong suppression and mitigation effects.

TABLE I. THE RELATIVE REDUCTION OF MITIGATION FACTOR AND SUPPRESSION FACTOR
UNDER DIFFERENT RESPONSE LEVELS OF 14 EUROPEAN COUNTRIES

| Response | Representative Measures | Impact of measures $R_t$ relative reduction | Suppression effect $D_t$ relative reduction | Mitigation effect $p_t$ relative reduction |
|---|---|---|---|---|
| **Level 0** **Minimal response** | No mandatory restrictions | 0 | 0 | 0 |
| **Level 1** **Soft response** | Closing schools, International travel controls. | 35% CI: [25%, 45%] | 22% CI: [17%, 27%] | 29% CI: [18%, 38%] |
| **Level 2** **Strong response** | Cancel public events, Restrictions on gathering, Restrictions on internal movement. | 60% CI: [54%, 65%] | 26% CI: [21%, 30%] | 56% CI: [50%, 61%] |
| **Level 3** | Close workplace, Close public transport, Stay-at-home requirements. | 71% CI: [68%, 74%] | 37% CI: [35%, 40%] | 67% CI: [64%, 70%] |
| **Level 4** **Emergent response** | | 74% CI: [71%, 77%] | 40% CI: [37%, 42%] | 70% CI: [66%, 73%] |

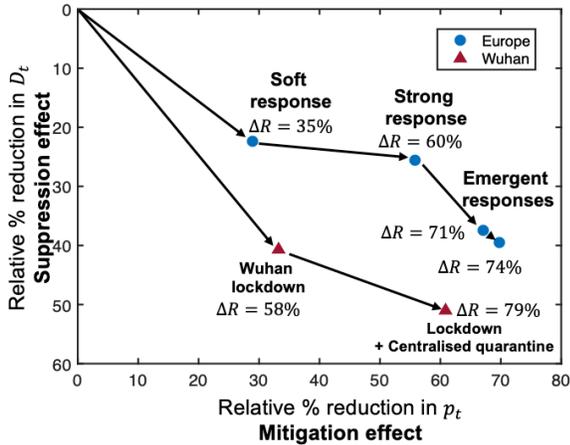

Fig. 9. The relative reduction of mitigation factor $p_t$ and suppression factor $D_t$ under different response levels compared to minimal response level.

### C. Resurgence risks in United States

We also used the proposed framework to estimate the epidemic evolution in different states of the United States. We observed that, as of the week ending May 31st, the averaged reproduction number $R_t$ in 30 states exceeds 1 (Figure 10). These could be related to the recent lift of government restrictions and alert us to take a close monitoring on the epidemic evolution.

At the time of preparing this paper (June 18th 2020), 29 out of the 30 states we alerted on 9th June 2020 have experienced an increased number of daily confirmed cases compared to that of May 31st, and 14 states have recorded all-time high after May 31st. When we prepare the final version in early August, this alarming prediction of a second wave outbreak is unfortunately proven true for all the states listed.

So far, the application of the framework to many countries and the retrospective impact analysis of intervention measures in European countries indicate the effectiveness of our approach in monitoring $R_t$. This can be further validated by predicting the evolution of $p_t, D_t$ and $R_t$ and projected infections in future study. Our current study has several limitations. Firstly, the reporting protocols and standards of confirmed cases, as well as the detection rates, vary among countries. However, as long as the reporting bias is consistent over time, the inference results of $p_t$, $D_t$ and $R_t$ should not be affected. We also note that the implementation of multiple intervention measures within a short interval makes it challenging to quantify the impact of a single measure which needs further statistical analysis.

### VII. CONCLUSIONS

In conclusion, we propose a comprehensive Bayesian updating approach to timely estimate parameters of COVID-19 epidemic models. The disease transmission dynamics is modelled by renewal equations with time-varying parameters. Instead of purely focusing on estimating instantaneous reproduction number $R_t$, we introduce two complementary parameters, the mitigation factor ($p_t$) and the suppression factor ($D_t$), to quantify intervention impacts at a finer granularity. A Bayesian updating scheme is adopted to dynamically infer model parameters. By monitoring and analyzing the evolution of the estimated parameters, impacts of intervention measures in different response levels can be quantitatively assessed. We have applied our method to European countries, the United States and Wuhan, and reveal the effects of interventions in these countries and the resurgence risk in the USA. Our work opens a promising venue to inform policy for better decision-making in response to a possible second-wave outbreak.

ACKNOWLEDGMENT

We express our sincere thanks to all members of the joint analysis team between Imperial College London, University of Cambridge and University of Kent and Hong Kong Baptist University. We thank Yuting Xing for helping collect epidemic data in Wuhan and the United States. We thank Siyao Wang and Liqun Wu for their efforts on developing a digital tracing app for validation and visualization.

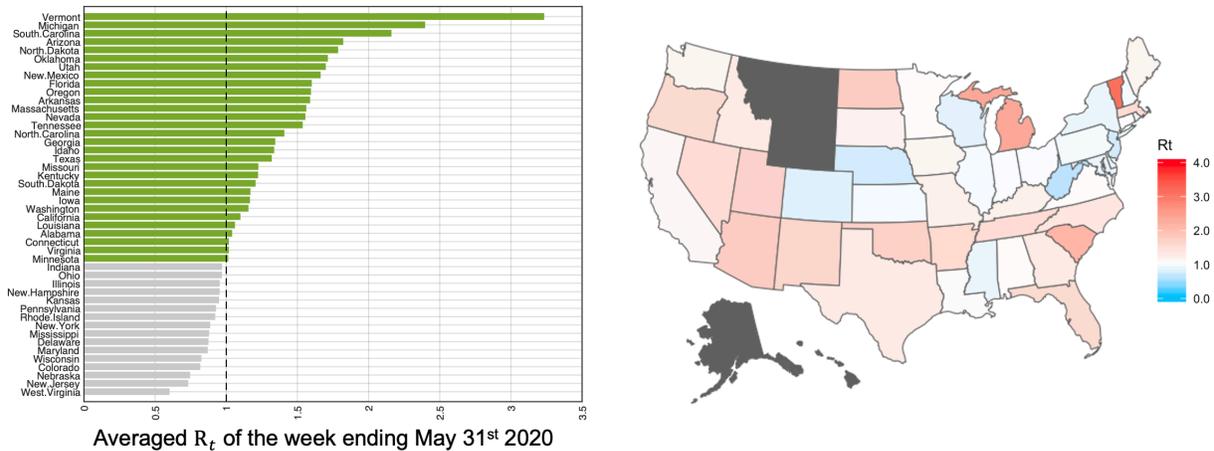

Fig. 10. The averaged $R_t$ values in different states of the United States. We report the result of averaged $R_t$ in the US during the week ending May 31st 2020, which is ranked by the averaged $R_t$ value (annotated with green if above 1, left). States with total confirmed cases less than 1,000 are excluded from the analysis.

## REFERENCES

[1] E. Dong, H. Du, and L. Gardner, "An interactive web-based dashboard to track COVID-19 in real time," *Lancet Infect. Dis.*, vol. 20, no. 5, pp. 533–534, May 2020.

[2] T. Hale, A. Petherick, T. Phillips, and S. Webster, "Variation in government responses to COVID-19," 2020.

[3] A. Pan *et al.*, "Association of Public Health Interventions With the Epidemiology of the COVID-19 Outbreak in Wuhan, China," *JAMA*, vol. 323, no. 19, p. 1915, May 2020.

[4] R. Li *et al.*, "Substantial undocumented infection facilitates the rapid dissemination of novel coronavirus (SARS-CoV2)," *Science (80-. ).*, vol. 3221, no. March, p. eabb3221, 2020.

[5] L. Ferretti *et al.*, "Quantifying SARS-CoV-2 transmission suggests epidemic control with digital contact tracing.," *Science*, vol. 6936, no. March, pp. 1–13, 2020.

[6] E. Vynnycky and R. White, *An introduction to infectious disease modelling*. OUP oxford, 2010.

[7] N. C. Grassly and C. Fraser, "Mathematical models of infectious disease transmission," *Nat. Rev. Microbiol.*, vol. 6, no. 6, pp. 477–487, 2008.

[8] C. Fraser, "Estimating individual and household reproduction numbers in an emerging epidemic," *PLoS One*, vol. 2, no. 8, 2007.

[9] J. Ma and D. J. D. Earn, "Generality of the final size formula for an epidemic of a newly invading infectious disease," *Bull. Math. Biol.*, vol. 68, no. 3, pp. 679–702, 2006.

[10] K. Leung, J. T. Wu, D. Liu, and G. M. Leung, "First-wave COVID-19 transmissibility and severity in China outside Hubei after control measures, and second-wave scenario planning: a modelling impact assessment," *Lancet*, vol. 395, no. 10233, pp. 1382–1393, Apr. 2020.

[11] S. Flaxman *et al.*, "Estimating the effects of non-pharmaceutical interventions on COVID-19 in Europe," *Nature*, pp. 1–5, 2020.

[12] R. N. Thompson *et al.*, "Improved inference of time-varying reproduction numbers during infectious disease outbreaks," *Epidemics*, vol. 29, no. August, 2019.

[13] A. Cori, N. M. Ferguson, C. Fraser, and S. Cauchemez, "A new framework and software to estimate time-varying reproduction numbers during epidemics," *Am. J. Epidemiol.*, vol. 178, no. 9, pp. 1505–1512, 2013.

[14] J. Wallinga and P. Teunis, "Different epidemic curves for severe acute respiratory syndrome reveal similar impacts of control measures," *Am. J. Epidemiol.*, vol. 160, no. 6, pp. 509–516, 2004.

[15] D. Adam, "A guide to R-the pandemic's misunderstood metric.," *Nature*, vol. 583, no. 7816, pp. 346–348, 2020.

[16] N. Imai, I. Dorigatti, A. Cori, C. Donnelly, S. Riley, and N. Ferguson, "Report 2: Estimating the potential total number of novel Coronavirus cases in Wuhan City, China," 2020.

[17] Q. Li *et al.*, "Early transmission dynamics in Wuhan, China, of novel coronavirus-infected pneumonia," *N. Engl. J. Med.*, vol. 382, no. 13, pp. 1199–1207, 2020.

[18] J. T. Wu, K. Leung, and G. M. Leung, "Nowcasting and forecasting the potential domestic and international spread of the 2019-nCoV outbreak originating in Wuhan, China: a modelling study," *Lancet*, vol. 395, no. 10225, pp. 689–697, 2020.

[19] P. Nouvellet *et al.*, "Report 26: Reduction in mobility and COVID-19 transmission."

[20] M. Asch, M. Bocquet, and M. Nodet, *Data assimilation: methods, algorithms, and applications*. 2016.

[21] Z. Chen, "Bayesian filtering: From Kalman filters to particle filters, and beyond," *Statistics (Ber).*, vol. 182, no. 1, pp. 1–69, 2003.

[22] C. J. Rhodes and T. D. Hollingsworth, "Variational data assimilation with epidemic models," *J. Theor. Biol.*, vol. 258, no. 4, pp. 591–602, 2009.

[23] L. M. A. Bettencourt and R. M. Ribeiro, "Real time bayesian estimation of the epidemic potential of emerging infectious diseases," *PLoS One*, vol. 3, no. 5, p. e2185, 2008.

[24] L. Cobb, A. Krishnamurthy, J. Mandel, and J. D. Beezley, "Bayesian tracking of emerging epidemics using ensemble optimal statistical interpolation," *Spat. Spatiotemporal. Epidemiol.*, vol. 10, pp. 39–48, 2014.

[25] E. Goldstein, J. Dushoff, M. Junling, J. B. Plotkin, D. J. D. Earn, and M. Lipsitch, "Reconstructing influenza incidence by deconvolution of daily mortality time series," *Proc. Natl. Acad. Sci. U. S. A.*, vol. 106, no. 51, pp. 21825–21829, 2009.

[26] C. Mark, C. Metzner, L. Lautscham, P. L. Strissel, R. Strick, and B. Fabry, "Bayesian model selection for complex dynamic systems," *Nat. Commun.*, vol. 9, no. 1, 2018.

[27] A. Agresti, *An introduction to categorical data analysis*. John Wiley & Sons, 2018.